EUV Irradiance Inputs to Thermospheric Density Models: Open Issues and Path Forward

A. Vourlidas[1,2] and S. Bruinsma[3]

[1]The Johns Hopkins University Applied Physics Laboratory, Laurel, MD, USA

[2] Also at IAASARS, Observatory of Athens, Athens, Greece

[3]CNES- Space Geodesy Office, Toulouse, France

Corresponding author: Angelos Vourlidas (angelos.vourlidas@jhuapl.edu)

**Key Points:**

- We review the use and limitations of EUV irradiance measurements for thermospheric density modeling.

- Indices based on EUV irradiance measurements are the best inputs for thermospheric modeling.

- We propose a set of recommendations for moving forward with improved thermospheric density modeling

**Abstract**

One of the objectives of the NASA LWS Institute on 'Nowcasting of Atmospheric Drag for LEO Spacecraft' was to investigate whether and how to increase the accuracy of atmospheric drag models by improving the quality of the solar forcing inputs, namely Extreme Ultraviolet (EUV) irradiance information. In this focused review, we examine the status of and issues with EUV measurements and proxies, discuss recent promising developments, and suggest a number of ways to improve the reliability, availability, and forecast accuracy of EUV measurements in the next solar cycle.

**1 Introduction**

Solar variability influences human society in many ways, from long-term climatic changes to telecommunications to the longevity of spacecraft. Of particular concern, here, are the effects of solar variability on the thermosphere (90 – 600 km altitude) where many spacecraft, including the International Space Station, reside. The main solar thermospheric drivers are (1) the extreme ultraviolet (EUV) radiant flux per unit area (irradiance) at wavelengths below ~200 nm and, (2) intermittent solar wind inputs from Coronal Mass Ejections (CMEs) and High Speed Streams (HSS) (Chen et al., 2012; McGranaghan et al., 2014). While the EUV irradiance varies on timescales from minutes to years, CMEs and HSSs are short time scale phenomena of the order of a few hours to days. Direct EUV emissions constitute the main source of heating of the dayside upper atmosphere (e.g. Lilensten et al., 2008), although the largest heating events are due to solar wind via Joule heating on very rare occasions (Knipp et al., 2004). Variable heating of the thermosphere causes changes in the density. This in turn affects the drag experienced by orbiting spacecraft, thereby altering their orbital elements--- notable, their semi-major axis (i.e. orbit altitude). Therefore, understanding and predicting variations in irradiance are crucial for improving the thermosphere density forecast, which plays a major role in collision avoidance, orbit and reentry predictions at heights above 90 km (see Lilensten et al. (2008) for a more general review of the Space Weather effects of EUV irradiance). Consequently, solar irradiance and its variations were a central issue in the discussions within the NASA Living With a Star (LWS) Institute on 'Nowcasting of Atmospheric Drag for LEO Spacecraft' convened in 2016-2017 to investigate ways to increase the accuracy of atmospheric drag models. This work is one of a series of papers reporting on the discussions within the Institute. As the creation of this LWS institute attests, the spatial and temporal variation of thermospheric density is of key importance for the scientific and operational communities and is a major area of research with several extensive and recent reviews (e.g., Qian & Solomon, 2012; Emmert, 2015; Liu et al., 2017).

To add value to this body of work, we focus our review on the solar drivers and proxies of solar drivers of thermospheric density and atmospheric drag changes, and discuss the open issues and path forward for improving both their forecast accuracy and time horizon for operational purposes. We do not specifically address any issues relating to the improvement of historical products and their re-analysis for research purposes. The selection of a proxy is not trivial for modelers. It requires choosing between using all available density data or using a better proxy. Our paper is organized as follows. In Section2, we introduce the operational requirements on the EUV irradiance measurements and their forecast accuracy, discuss how these measurements or their proxies are used to model the thermospheric response, and outline their shortcomings. The forecast accuracy of EUV lines and spectral bands (e.g. HeII at 30.4 nm) and proxies are discussed in Section 3. In Section 4, we suggest a path forward for improving atmospheric drag predictions based on EUV irradiance inputs.

**2 EUV Irradiance and its Proxies**

The Sun drives the density in the thermosphere via direct and indirect heating of this layer. Absorption of EUV (10 – 120 nm) and UV (120 – 200 nm) radiation accounts for ~80% of the energy input into the thermosphere. This raises the thermospheric temperature leading to expansion and contraction as the passage of active region across the solar disk cause the EUV irradiance to rise or dip. At the same time, as solar radiation at longer wavelengths heats the lower atmosphere, part of the energy dissipates upwards, as thermal tides, affecting the thermospheric temperature structure (e.g. Liu, 2016). Both solar forcing components lead to density perturbations across the globe as the system tries to adjust. Predicting density variations is of paramount importance for orbital dynamics but the task is complicated by the multiscale nature of the EUV variations and by instrumental constraints of EUV monitoring from space. We proceed to discuss the various approaches and issues in the measurements of EUV irradiance and its use in thermospheric models.

2.1 Measurement Requirements for the EUV Irradiance

To frame the issues surrounding EUV irradiance observations, it is helpful to understand the measurement (e.g. accuracy, spectral coverage) and forecast (e.g. lead time) requirements on the EUV observations for operational thermospheric modeling. In other words, how good do these measurements need to be and how far in the future should their forecast extend?

In the U.S., many of the operational requirements for terrestrial and space weather observations are captured within the National Polar-Orbiting Operational Environmental Satellite System (NPOESS) Integrated Operational Requirements Document (IORD) I (IPO 1996). The document specifies (req. 4.1.6.7.15) that the EUV measurements should be obtained over 5-130 nm in 4 channels, with an accuracy of greater than $\pm 10^{-4}$ W/m$^2$ or $\pm 20\%$, and cadence of 5 hours. These requirements were used to guide the development of the EUV sensors aboard the National Oceanic and Atmospheric Administration's (NOAA) Geostationary Operational Environmental Satellite (GOES) satellites, the first of which was launched on GOES 13 in 2006. The requirements have evolved over the years and the current ones call for (R. Viereck, personal communication):

- a complete EUV spectrum (5 – 120 nm) with 5 nm resolution,
- at 30-sec cadence,
- with 30-sec latency.
- Accuracy of 20% needs to be maintained for over 10 years of satellite life.

The GOES-16 Extreme Ultraviolet and X-ray Irradiance Sensors (EXIS), launched in 2016, will be the first ones to fulfill these requirements. To put these requirements in context, we can examine the effect of irradiance measurement accuracy on collision avoidance calculations from Emmert et al. (2014).

They developed equations relating the uncertainty in the EUV irradiance for a given conjunction rate. The latter is defined as the number of objects (usually debris) intersecting an ellipsoidal volume around the object of interest (i.e., spacecraft) during its orbit. Operationally, the Air Force Joint Space Operations Center (JSpOC) uses oversized volumes (e.g., 44 km in the in-track direction at ~400 km altitude) that do not vary with solar activity. To investigate the effect of solar activity, Emmert et al. (2014) depart from the operational definition above and assume that the orbit of interest is known and the ellipsoidal volume of the secondary objects grows with solar activity. Then, the uncertainty estimate depends on the desired daily conjunction rate at a particular activity level, approximated by the solar radio flux at 10.7cm ($F_{10.7}$ index; see Section 2.5, for details), an EUV irradiance *forecast* uncertainty of <25% would be sufficient to keep the daily number of conjunctions below 5 for a 3-day forecast. Note that these estimates are not the same as the probability of collision ($P_c$) that is the operational quantity used for CARA. Only the number of conjunctions with $P_c$> 1e-4 should be considered for operational requirements. At any rate, the Emmert et al. (2014) attempts to provide some quantification on the solar activity effects on CARA and suggests that the EUV measurements themselves have to be accurate to much better than 25% (1-3%, similar to $F_{10.7}$) to provide operational benefit; a 25% error in the solar activity typically causes density errors of 10-100%, depending on altitude and solar activity level. An additional complication arises in case of thermosphere model development and the ingestion of composition measurements taken in the '70s and early '80s---the accuracy and long-term stability of the index over tens of years must be at the 1-3% level in order to reproduce density consistently both almost fifty years ago and at the present date. This requirement is met only by radiometric measurements taken from the ground ($F_{10.7}$ and $F_{30}$, see Section 2.5), also when comparing only over the 20 years, i.e.

since the advent of Solar EUV monitor (SEM) data aboard the Solar and Heliospheric Observatory (SOHO) mission. The in-orbit absolute calibration of the EUV measurements may therefore be an issue, when one attempts to construct a consistent (composite) time series that contains long data gaps or no overlap between consecutive satellite missions.

Turning to the lead time for accurate EUV forecasts, the requirements can be derived from the requirements for the neutral density specification and the practices for Collision Avoidance and Risk Analysis (CARA). The Air Force requires accurate density forecasts over the next 72 hours. CARA orbit predictions range from a 7-day forecast for Low Earth Orbit (LEO) objects to 10-day forecast for geosynchronous objects (Newman, 2010). Forecast for longer horizons also exist (Section 3), driven by the need for long-range planning (up to 45 days out) or lifetime estimates (years to several solar cycles). It seems, therefore, that the most important lead-time requirements, for our purposes here, are 1-10 day EUV irradiance forecasts. We discuss their accuracy in Section 3.

### 2.2 EUV Irradiance Measurements.

EUV irradiance has been measured since 1960 with the SOLRAD and Orbiting Solar Observatory (OSO) missions, followed by AEROS and the Atmospheric Explorer (AE) in the 1970s. After a period of very sparse coverage in the 1980s – 1990s, (Figure 2 in Woods et al., 2005) regular EUV coverage returned with the SOHO/SEM continuously operating since 1995. The measurements covered very different intervals, mostly the short range in the EUV and Soft X-rays before 1970, later extending into the EUV/UV (see Table 1 in Woods et al., 2004). The instruments are variously high or medium resolution spectrometers or broad-band photodiodes or even imaging instruments (in the case of Yohkoh, for example). Besides the non-uniformity of spectral coverage and resolution, the early solar EUV irradiance measurements, reviewed by Woods et al. (2004), were plagued by intermittent coverage (e.g. 3-5 days of observations per week, requiring interpolation for the missing days, and therefore loss of accuracy) and cross-calibration effects.

In 2002, the Solar EUV Experiment (SEE) aboard the Thermospheric Ionospheric Mesospheric Energy and Dynamics (TIMED) mission (Woods et al., 2005) established synoptic coverage with regular calibration rocket underflights. The coverage has now been improved with the Extreme Ultraviolet Variability Experiment (EVE; Woods et al., 2012) aboard the Solar Dynamics Observatory (SDO). EVE provides measurements from 0.1 to 105 nm at 0.1 nm spectral resolution and 10-sec cadence since 2010. Unfortunately, a power anomaly disabled the Multiple Extreme ultraviolet Grating Spectrograph (MEGS)-A instrument on May 27, 2014 and hence EUV and soft X-ray spectra below 37 nm are no longer available. The Extreme Ultraviolet Spectrophotometer (ESP), an expanded version of the SOHO/SEM instrument, fills partially the gap with 5 broadband channels between 0.1 -37 nm, including the 27-35 nm band from SOHO/SEM. The remaining spectrograph, MEGS-B (35-105 nm range with 0.1 nm spectral resolution) is still operational but faces unexpected and rapid degradation so it is operating at a 10-sec cadence for approximately 3 hours/day to prolong its life.

2.3 Calibration strategies for EUV irradiance measurements

The main source of uncertainty with EUV irradiance measurements is the *inadequate* monitoring of their absolute calibration as a function of time. This is an issue for all EUV instruments, whether they are imagers, spectrometers or radiometers because they all suffer significant degradation and sensitivity loss making it difficult to establish the reliability of long-term trends.

The degradation stems from four sources mainly (e.g. Cessateur et al., 2012):
1. contamination of the entrance filters,
2. change in the spectral responsivity of the interference filters,
3. stray light effects from pinholes on the filters,
4. and detector degradation.

Entrance filter contamination can lead to very sharp drops in the instrument sensitivity in the early phases of the mission and is attributed to polymerization of hydrocarbon contaminants. Although this contamination source is understood and controlled during the development of modern missions, instrument sensitivity loss still occurs suggesting that other mechanisms, possibly oxidation, may be at play. However, there have been very few studies of these effects on UV/EUV filters, as Cessateur et al. (2012) point out. The changes in the spectral responsivity may arise from the contamination of the entrance filter or structural changes due to the solar UV radiation or both. We currently lack the systematic information needed to understand this effect. Pinhole effects are difficult to guard against for the required thicknesses of the filters. They could be mitigated with the inclusion of backup filters albeit increasing the complexity of the instruments. The detector degradation is usually a long-term trend and is a function of the technology used in the specific instrument. Microchannel plate (MCP) detectors, for example, suffer from irreversible gain losses due to radiation. CCDs, on the other hand, can recover most of their gain via thermal cycling (annealing). Diamond-based or Active Pixel Sensor detectors are much less susceptible to radiation effects (e.g. BenMoussa et al., 2013).

The need for in-orbit calibration and cross-calibration strategies has been long understood (Pauluhn, Huber & von Steiger, 2002). In the case of EUV radiometers, the TIMED/SEE instrument was the first experiment to adopt a detailed pre-flight calibration and regular underflights with calibration sounding rockets to monitor instrument performance. At launch, TIMED/SEE had an absolute irradiance uncertainty of 10%-20%, based on wavelength, with the largest uncertainties for wavelengths below 27 nm. The short wavelength observations are provided by the XUV Photometer System (XPS) broadband diodes and use a model to reconstruct the spectral irradiance in this range, so larger uncertainties are expected anyway. The uncertainty increases by a few % per year but rocket underflights every two years monitor the degradation (Figure 1, Lean et al., 2011). The rockets carry a copy of the SEE instrument which is calibrated at a National Institute of Standards and Technology (NIST) facility both before and after the rocket flight. These measurements are used to update the SEE irradiance products but the reliance on underflights also means that such updates may be unavailable when the launches fail, as in 2015. That failure impacted the accuracy of the measurements in 2012-2015 (level 11) until the next successful flight in 2016 (version 12). Degradation updates in the EUV irradiance data products propagate downstream to updates in proxies that depend on them, for example $S_{10.7}$ (defined in Section 2.5). However, it may not always be clear to the end user when or why these updates occur. We revisit this issue in Section 2.4.

Taking advantage of the TIMED/SEE experience, the SDO/EVE instrument implements a more rigorous calibration strategy described in Didkovsky et al. (2012) for the ESP, and Hock et al. (2012) for the MEGS. The data products and the concept of operations of EVE are described in Woods et al. (2010). The project involves yearly calibration rocket flights, the latest occurring in August 2016. It is, therefore, apparent, that even after 40 years of space-based observations, *our knowledge of sensor and filter degradation is limited and more effort should be invested in understanding EUV degradation and in finding ways to avoid it* (statements in italics refer to the 'Path Forward' consideration in Section 4).

### 2.4 Temporal variation of the EUV irradiance

The UV/EUV radiation arises from different heights of the solar atmosphere with UV originating in the upper chromosphere from plasmas at temperatures $8 \times 10^3$ to about $10^5$ K and most of the EUV coming from higher up in the corona where the emitting plasmas are at million degree temperatures. The magnetic field begins to dominate over the plasma pressure in these heights, resulting in very short Alfvén time scales of seconds to minutes. Consequently, magnetic energy can be released explosively to heat up the chromospheric and coronal plasmas in a matter of seconds (in flares) to minutes and hours (in CMEs) and giving rise to EUV (and soft X-ray, for large flares) variations at the same time scales (Woods, Knopp & Chamberlin, 2006). Daily variations arise from the emergence of new active regions (ARs) on the solar disk or the appearance of developed AR loop systems at the limb. The passage of ARs across the disk creates variations at the weekly to bi-weekly level and eventually 27-day variations are driven by the solar rotation rate which brings these ARs back. Timescales of months to 1-2 years are correlated with active longitudes--extended areas with persistent emergence of AR systems--while yearly to decadal timescales are correlated with the solar cycle evolution.

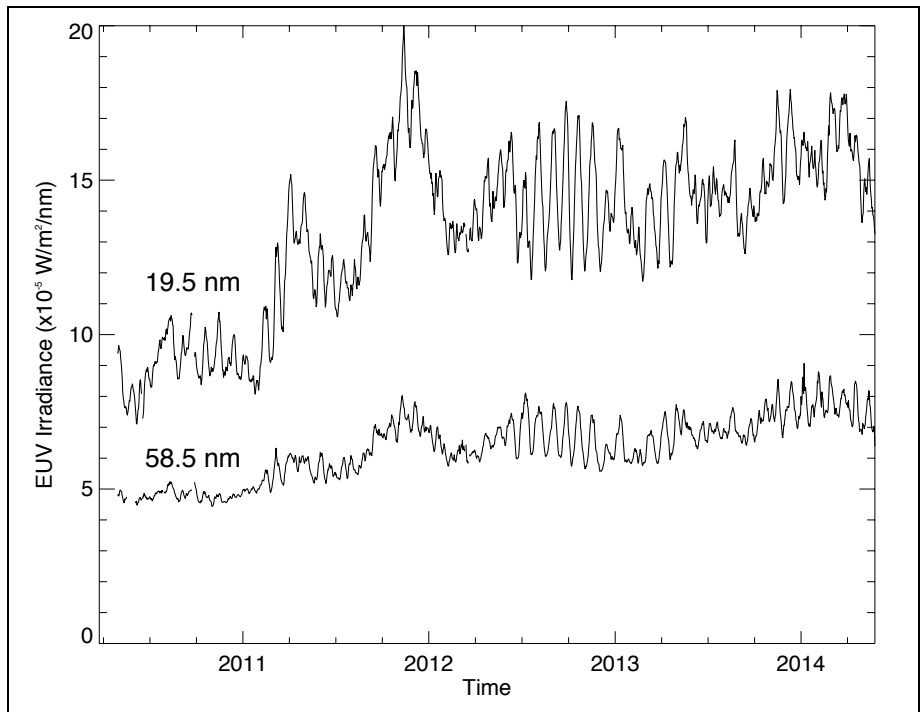

**Figure 1**. Time series of the spectral solar irradiance at 58.5 (He I) and 19.5 nm (Fe XII) from the EVE instrument aboard the SDO spacecraft. The wavelengths represents solar emission from plasmas at around 80,000K and 1.4 MK, respectively. The plots are based on L3 data (and their calibration) as downloaded from the EVE online repository (on Dec 31, 2017) and may contain instrumental artifacts (http://lasp.colorado.edu/lisird/tss/sdo_eve_ssi_1nm_l3.csv).

The EUV irradiance varies significantly in all time scales (and across wavelength) as can been in Figure 1, which shows daily averages of EUV irradiance from SDO/EVE at 58.5 and 19.5 nm for the rising and maximum phase of solar cycle 24. The wavelengths are chosen because of their proximity to the He I (58.4 nm) and Fe XII (19.5 nm) spectral lines that arise from cool (~0.08 MK) and hot (1.4 MK) coronal material, respectively. The flare EUV radiative output can easily exceed 3-4 times the background level (e.g., Chamberlin, Milligan & Woods, 2012) while reappearing ARs produce fluctuations of the order of 25% at 58.5 nm and 40% at 19.5 nm, for example (second half of 2012 in Figure 1). Clearly, these levels are highly wavelength-dependent and can vary by an order of magnitude across the whole EUV range. A factor of two change is typical over the course of a solar cycle (Figure 1; Lean et al., 2011).

Flare-induced irradiance changes take about 1-2 hours to alter the thermospheric density but the changes are most often imperceptible; only extreme X-class flares produce density enhancements of the order of tens of percent (Qian et al., 2010; 2011) The largest perturbations (50-60%) were observed on the CHAMP and GRACE spacecraft for the 28 October 2003 X17.2 flare (Sutton et al., 2006). Even for X-class flares, most perturbations are of the order of a few percent (Le et al., 2012; Pawlowski & Ridley, 2008). X-class flares are rare, varying from 0 to 20 per year, and those that cause measurable density perturbations, (X-class

> X9), even more so. With an occurrence rate of ~2 10$^{-3}$ /day (Winter et al., 2016), we expect only a handful (~8) of X9+ events per solar cycle. This may be a reason for the relative scarcity of studies on the thermospheric response to flares. Most of these studies are summarized by Emmert (2015). Flare-driven thermospheric modeling seems to hold relatively little operational forecasting value given: (1) the lack of any reliable methods for flare predictions, in the near- to mid-term, at least, (Guerra et al, 2015), (2) the insensitivity of the thermosphere to < X-class flares, which constitute the grand majority of flare events, and (3) the short thermospheric reaction span (1-2 hours). The low flaring activity in solar cycle 24, and a similar level anticipated for cycle 25, provide even less urgency for operational advances in this area.

### 2.5 EUV Proxies

The interrupted record of solar EUV irradiance and the degradation of the measurements, discussed in 2.2, present obstacles for using the solar measurements in operational settings for orbit prediction. The requirements of the operational community for timely, reliable and easily interpretable irradiance information have led to the adoption of proxies for constructing irradiance variability models and for driving thermospheric models. The proxies need to satisfy three criteria: (1) real-time and uninterrupted availability, (2) frequent, consistent and accurate calibration, and (3) as representative as possible of the actual solar EUV irradiance on all time scales (days to years). The cadence of most proxies is daily, and (linearly) interpolated values are used in thermosphere models. The day-to-day variability in EUV can be quite large (10-20%), and sub-daily cadence of the proxies would diminish (by an as yet unknown amount) the error due to the interpolation.

The most popular proxy is the daily $F_{10.7}$ index (Tapping, 2013), corresponding to the radio flux at 10.7 cm (2.8 GHz), followed by the Mg II index (Heath and Schlesinger, 1986), which is the core-to-wing ratio of the Mg II line forming at 280 nm. The Mg II index has been available since 1978 from space-based measurements on several consecutive satellites starting with Nimbus-7, but not uninterruptedly. The index depends on the spectral resolution of the measurements and hence requires cross-calibration to produce a coherent long-term dataset (e.g. Viereck et al., 2004), i.e., a composite. The $F_{10.7}$ index, in contrast, has been available since 1947 from the Dominion Radio Observatory and represents a very stable irradiance proxy thanks to methodical daily calibrations. The observations are almost always available. The interruptions amount to a total of a few weeks scattered over the index's entire existence. For these reasons, it is also the most widely-used index, even though it does not completely represent all sources of solar EUV irradiance (Lean et al., 2011).

Indeed, the Mg II index outperforms $F_{10.7}$, as a proxy to the UV (> 27 nm) where the emission arises from cooler chromospheric plasmas (Lean et al., 2009). The $F_{10.7}$ index performs better at shorter wavelengths since the 10.7 cm emission is mainly due to gyroresonance emission, and thermal 'free-free' emission at solar minimum, from electrons at million-degree plasmas (e.g. Vourlidas et al., 1997, Dudok de Wit et al., 2014; Schonfeld et al., 2015). It is an obvious simplification to represent the complexity of solar atmospheric emission with an effectively two-component model (i.e., trend and variations on solar rotation and shorter time scales). Naturally, other indices and proxies have appeared in the literature to bridge this gap

(see Tables 1-2 and Figure 2 in Tobiska et al. (2008) for a list of indices and proxies and their spectral coverage).

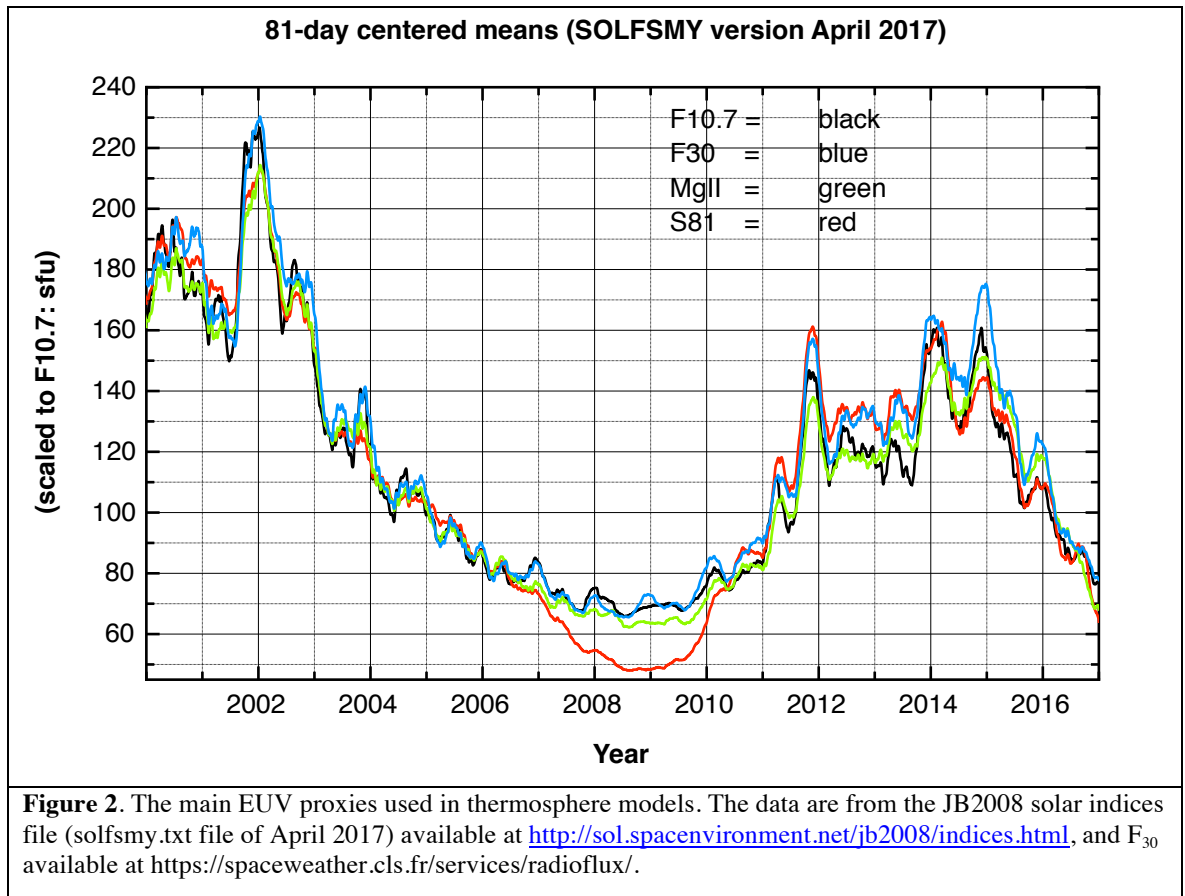

**Figure 2**. The main EUV proxies used in thermosphere models. The data are from the JB2008 solar indices file (solfsmy.txt file of April 2017) available at http://sol.spacenvironment.net/jb2008/indices.html, and $F_{30}$ available at https://spaceweather.cls.fr/services/radioflux/.

Radio observations at other centimetric wavelengths have recently been shown to offer better performance than $F_{10.7}$ (Dudok de Wit et al., 2014; Dudok de Wit and Bruinsma, 2017), notably the 30 cm flux ($F_{30}$). The Japanese radiopolarimetric observations, first by the Toyokawa and then Nobeyama observatories, started in 1951 and have acquired over 60 years of daily observations with high radiometric stability and very few interruptions (observed and predicted values: https://spaceweather.cls.fr/services/radioflux/). Because of its better performance, ongoing availability over the entire space age, and robustness (ground measurement), $F_{30}$ was selected as EUV proxy in replacement of $F_{10.7}$ in the Drag Temperature Model (DTM2013; Bruinsma, 2015). $S_{10.7}$, originally based on the 26-34 nm integrated flux from the SEM radiometer on SOHO (Judge et al., 1998), available since 1997, and representing a proxy of the He II 30.4 nm emission, is used in the operational model JB2008 (Bowman et al., 2008). Figure 2 displays the main EUV proxies, all of which (except $F_{30}$) are used in the JB2008 model from 2000 through 2016. The large difference with S81(the 81-day average of $S_{10.7}$) from 2006-2011 is not due to solar variability, but due to a specific JB2008 model correction in order to absorb the effect of thermospheric cooling.

The question now arises which of these indices and proxies is the best for thermospheric modeling. Dudok de Wit & Bruinsma (2011) studied this question and found that the $S_{10.7}$, and generally the 26-34 nm integrated flux, is the best index for modeling the thermospheric density at 813 km altitude. They also found that multi-index combinations do not necessarily improve operational models and suggest that transfer function models can account for the time variable response of the thermosphere better than the static indices. Therefore, it is essential that instruments either measure the identical range (for continuity), or have sufficient resolution to accurately reconstruct it. This requirement excludes certain instruments for thermosphere modeling purposes (e.g., the Large Yield Radiometer (LYRA; Dominique et al., 2013) aboard the PROBA2 spacecraft). A crucial disadvantage of satellite EUV measurements and therefore also He II-based proxies is, besides calibration and cross-calibration issues due to different instruments, that the main composition measurements we have were taken by Dynamics Explorer-2 in the early '80s and the Atmosphere Explorer (AE-C, AE-D and AE-E) in the '70s. They can only be assimilated in a thermosphere model that uses $F_{10.7}$ or $F_{30}$.

Another approach is to cover the UV/EUV range by combinations of EUV measurements within discrete passbands (e.g. Dudok de Wit et al., 2008; Cessateur et al., 2011; Suess et al., 2016) or by using additional information, such as expanding the $F_{10.7}$ index with additional radio wavelengths that encompass emission from other layers of the solar atmosphere (Dudok de Wit et al., 2014) and, indirectly, serve the same purpose as the EUV measurements. Other researchers prefer to use spatially resolved images in the UV/EUV to estimate the contribution of different solar features (i.e. sunspots, plages, coronal holes) to the spectral irradiance via semi-empirical methods, i.e. differential emission measure (Warren et al., 2001) or image segmentation and synthetic spectrum (Haberreiter et al., 2014). Another, more recent approach relies on empirical relationships between photospheric magnetic flux and EUV emission (e.g., Henney et al. (2015), and references therein). All these approaches increase the physical understanding of the variations in solar EUV irradiance and improve the sophistication of irradiance inputs to the thermospheric models. Their inputs may also become more relevant operationally as the full disk EUV imaging will be available from GOES-R and the Henney et al., (2015) method relies on a photospheric magnetic flux transport model designed specifically for operational deployment. These approaches hold great promise for the future, especially those based on magnetic field observations that have many advantages over existing indices and proxies; namely, ready availability of photospheric magnetic field observations from ground and space, less susceptibility to instrument degradation, spatially resolved measurements, and better scientific understanding of magnetic field evolution (Warren & Emmert, 2014). But to be considered for operational deployment, the forecast accuracy of these approaches needs to be tested against the performance of the current operational inputs, such as $F_{10.7}$ or other proxies. For example, Warren et al. (2017) find that the Henney et al. (2012) method performs at a similar level to the (much simpler) $F_{10.7}$ linear forecast. As our knowledge on the evolution of solar magnetic fields improves so will such models. Therefore, while none of these approaches can be adopted *yet* by operational thermospheric density models, they should nevertheless be benchmarked against the operationally deployed models.

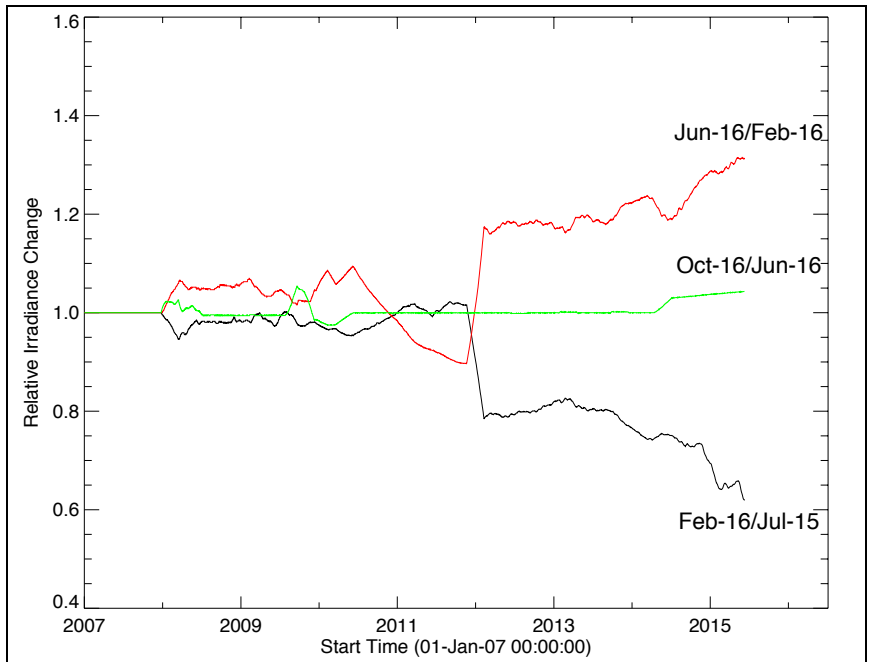

**Figure 3**. Relative changes in the 81-day averaged $S_{10}$ over four successive releases of the solar indices data. The data are from the JB2008 solar indices file (solsfmy.txt file) available at http://sol.spaceenvironment.net/jb2008/indices.html and represent ratios between solsfmy.txt files released on the dates indicated on the plots. Some of the changes correspond to updates to the irradiance calibrations but they are not documented in detail. The figure demonstrates the inconsistency of the same S10.7 dataset downloaded at different times. different versions of the same data can have discrepancies of up to 20% in their input irradiance values.

This body of work makes clear a rather obvious point: proxies or approaches based on the EUV measurements themselves are the best inputs for thermospheric modeling since they are a direct representation of the solar input rather than an indirect proxy, like $F_{10.7}$ or the photospheric magnetic field measurements. The flipside is that are limited by the instrumental effects and calibration strategy of the EUV instruments they are derived from. This makes it difficult to study the EUV irradiance variations in the past.

Even from 2002 onwards, with the EUV measurements provided from SEE and EVE, and well-planned calibration strategies, problems remain. A case in point, are changes in $S_{10.7}$ over the last couple of years due to updates on the EVE calibration. Space Environment Technologies maintains the JB2008 operational model and provides a version-controlled list of daily values for several indices, including $S_{10.7}$, $F_{10.7}$, $M_{10.7}$ etc., for use by the community. The last four versions of this file are plotted as ratio of the various files, in Figure 3. There are changes in February 2016 marked with a new version (4_2h) but their origin is not noted in the file. Changes in June 2016 (version 4_2g) seem to reverse the previous changes and are marked as empirical degradation corrections (likely an TIMED/SEE update). Another update was released on October 2016 with the same version but the origin of the, albeit small (< 5%), changes are not noted in the file. They are likely due to further updates to the instrument's calibration but the lack of documentation makes it risky to implement within an operational environment. *Such changes, therefore, need to be tightly controlled and documented, to provide true operational value*. Differences in $S_{10}$ are tens of percent over several years; this causes differences in the tens to hundred percent in density, depending on altitude and solar cycle phase, and the JB2008 model performance has to be recomputed for each proxy version. Also, tests become unrepeatable unless one archives the used proxy file (past version do not appear to be available online).

## 3 Forecast Accuracy

Besides the obvious need for accurate EUV measurements, accurate forecasting of EUV irradiance variations is also very important, especially for orbit and collision avoidance predictions (Section 2.1). The National Weather Service (NWS) Space Weather Prediction Center (SWPC) issues 1-3, 27, and 45-day forecasts only for the $F_{10.7}$ index. Predictions 30 days out are also available for $F_{30}$ (ftp://ftpsedr.cls.fr/pub/previsol/solarflux/forecast). There are no such freely-available forecasts for Mg II, $S_{10.7}$ or other EUV-related indices or proxies. Lean et al. (2009) showed that MgII-based forecasts are superior to $F_{10.7}$ ones for 1-10 days, using autoregressive techniques which themselves outperform forecasts based on persistence or climatology. Naturally, the errors increase as the forecast is extended into the future. Emmert et al. (2017) made a detailed study of error propagation to density models from EUV inputs. They show that it is straightforward to generate hourly-resolution EUV irradiance averages instead of the standard daily ones and find that simple linear extrapolations work equally well to the autoregressive ones. The extrapolations are generally within the 5% density uncertainty requirement (Section 2.1) for 2-3 day forecasts (Figure 3j in Emmert et al., 2017) but the 10-day EUV forecasts behave as Brownian motion and hence lead to time$^5$ increases for in-track errors. More recently, Warren et al. (2017) evaluated the performance of a linear extrapolation model for $F_{10.7}$ and showed that the forecast skill peaks at 6 days compared to climatology and persistence.

Hence, statistical methods remain the best option for time scales of weeks and longer, albeit with poor accuracy. The forecast accuracy for solar cycle or longer time scales is presently too low to be of use for satellite operators, as is demonstrated by Pesnell (2015), who compared solar cycle 24 predictions with the observed maximum. Instead of using a prediction, in most instances standard cycles (low, moderate, high activity) are used in mission design, for example, the ISO space environment standards (ISO 14222).

Therefore, *finding a way to improve the forecast reliability beyond 7 days in the future and hence constrain these EUV errors would be of great benefit for collision avoidance (primarily) and mission planning (secondarily) applications*. This will likely require better knowledge of the solar evolution in short time scales (i.e. the reappearance of active regions over a solar rotation is already included in the models). There has been significant progress on both the irradiance modeling of solar features (e.g. Domingo et al. (2014), and references therein) and in understanding the evolution of the magnetic field on the solar surface via flux transport (Arge et al., 2010; Upton & Hathaway, 2014). Since EUV variability is driven by magnetic field changes, forward modeling of the magnetic field offers a semi-empirical way to forecast EUV with very positive results (Henney et al., 2015). This approach may work for longer time scales, say, 1-2 solar rotations as flux transport models seems capable in reproducing the EUV flux, in He II 30.4 nm in this case, quite successfully (Ugarte-Urra et al., 2015). The largest discrepancies between predicted and actual EUV irradiance were due to flux emergence on the far-side of the Sun. Although far-side imaging from Earth-based magnetographs is now routinely available (Lindsey and Braun, (2017), see also http://farside.nso.edu and http://jsoc.standford.edu/data/farside), it is not sufficiently robust for operational thermospheric modeling. The technique can only detect regions above a certain EUV intensity as measured in STEREO images (Liewer et al., 2017) and requires 5-day averaging of the Doppler data to increase the seismic sensitivity.

Naturally, the most direct way to improve EUV forecasts for terrestrial Space Weather are direct measurements of the EUV irradiance and photospheric magnetic field over the eastern solar limb. This information would immediately improve the robustness of 1-3 day EUV forecasts (required for CARA and density predictions) based on autoregressive or linear transformation models as in Lean et al. (2009) and improve the longer-term forecasts (7-10-day for orbit determination) based on forward modeling of the photospheric magnetic flux evolutions (Henney et al., 2015, Ugarte-Urra et al., 2015). These observations can be readily provided by a platform at the Lagrangian $L_5$ point as suggested by Vourlidas (2015) although the current operational concept for this mission does not include EUV radiometers (Trichas et al., 2015).

**4 Path Forward for EUV Irradiance Inputs**

In our discussions within the LWS Institute we identified three high-level issues with EUV irradiance measurements as inputs to thermospheric models: (1) the calibration and monitoring/in-flight calibration of instrument degradation remains the main problem for the adoption of direct EUV measurements as inputs to thermospheric density models, (2) the user-desired forecast accuracy of a few % (nominally 5% to match the requirement on neutral density) is limited to less than the required 3 days, (3) the physical relationship between EUV and thermospheric density variation is complex and remains an empirical rather than a physics-based parametrization.

The quality of the thermospheric forecasting could be improved if we could address the shortcomings in the treatment of the solar forcing input. In this review, we touched on four specific themes: (1) individual proxies or indices provide an incomplete characterization of the solar EUV output, and their performance is proportional to lead-time reducing to climatology after about 7 days; (2) changes and updates to the indices used in operational settings need to be documented and controlled; (3) there is need for a clear connection between a given proxy, the solar input it represents, and the thermospheric altitude or time-scale variation it is more appropriate for; (4) there needs to be an improvement in the connection between proxies or indices and the physical processes in the Sun that give rise to the EUV emission.

As a final product of our input to the LWS Institute, we have compiled a list of recommendations to address these issues. They are not in priority order but rather in order of ease of implementation. We hope that the list serves as a useful guide for resource allocation within the LWS program. Our recommendations are:

1. Compile or collect (if such requirements exist) a set of requirements on EUV irradiance measurement accuracy, including forecast accuracy for specific horizons (e.g. 1-3 days for density specification or 7-days for LEO object CARA) specific to the needs of operational thermospheric modeling.
2. Develop a set of benchmarks for the consistent testing of the performance of new indices, proxies, and EUV reference spectra.

3. Develop a 'rules of the road' for the development and maintenance of indices/proxies/spectra used in thermospheric models. The rules should describe how and when these inputs are updated, the calibration and assumptions that have gone into them and how this information should be disseminated in the user community. The approach and lessons-learned during the recent recalibration of the Solar Sunspot Number (Clette et al., (2016), and references therein) may be a useful guide.
4. Maintain EUV irradiance monitoring and ensure sufficient overlap between successive missions to allow for cross-calibration (e.g. between SDO/EVE and GOES-R/EXIS). Future measurements should be consistent with the requirements in Section 2.1 (follow GOES-R/EXIS, in other words), if a spectrometer is considered. When this is not possible, the absolute minimum EUV monitoring that can be useful for thermospheric modeling should include at least three spectral bands: HeII (26-34 nm) since the SOHO/SEM band is the most important to maintain for the semi-empirical thermosphere models, MgII (280 nm) since the MgII index is more accurate than $F_{10.7}$ as discussed previously, and Lyman alpha (118-122 nm) since it is the dominant line in the solar spectrum with effects across a large part of the terrestrial atmosphere. The combination of these spectral bands can provide quite accurate irradiance models (in lieu of the full spectrum) as Cessateur et al. (2011) and Suess et al. (2016) have demonstrated.
5. Address instrument degradation in space with technology development for current designs (more resilient entrance and interference filter, for example), support investigation into its causes (carbonization vs. oxidation) through the Heliophysics Technology and Instrument Development for Science (HTIDeS) program, and encourage the development of new instrument designs resilient to degradation and with in-flight calibration capabilities.
6. Support research on physics-based models of solar EUV irradiance and their transition to operational status.
7. Investigate how the forecast accuracy of EUV irradiance is improved via observations over the eastern solar limb from off Sun-Earth viewpoints (e.g. from the Lagrangian $L_5$ point). The degree of improvement, the nature of the sensors (EUV radiometers, EUV imagers, photospheric magnetic field observations) and the range of angular distances most beneficial to the operational community.

In closing, we have reviewed the work on EUV irradiance predictions vis-à-vis thermospheric density model with an emphasis on the issues surrounding the forecasting of the EUV irradiance. The last twenty years or so have witnessed great strides in our understanding of the sources and variability of EUV irradiance thanks to improvements on multiple fronts. Long, uninterrupted spectral irradiance observations from TIMED/SEE, SOHO/SEM, SDO/EVE, and soon, GOES/EXIS are bow available (Sec. 2.2). Instrumental degradation and the need for ground and in-flight monitoring of instrument throughput are now better understood and implemented (Sec. 2.3). The availability of spatially resolved EUV images and photospheric magnetograms from SOHO and SDO is inspiring novel approaches in modeling EUV irradiance and understanding the advantages and shortcomings of the various proxies (Sec 2.5). The field is vibrant, with multiple teams across the world, inventing and accessing models based variously on EUV measurements, indices, solar atmospheric modeling or combinations of these. Most models can outperform simplistic statistical model (such as persistence or recurrence) out to 7-day forecasts, at least and we have every reason to believe that their accuracy will improve as our physical understanding of the connection between EUV emission and magnetic field improves and as reasources are allocated to address the 'bottlenecks' identified during the LWS Institute.

**Acknowledgments**

The authors are not aware of any real or perceived financial conflicts of interests relating to this manuscript. AV was partially supported by the NASA LWS program through ROSES NNH13ZDA001N. We gratefully thank the referees and the editor for helpful comments that have improved the manuscript substantially. We also wish to thank H. Warren, L. Paxton, and R. Viereck for helpful discussions. We thank the SEE, EVE, and SEM teams for making their data publicly available and K. Tobiska & D. Bower for the $S_{10}$ data and related discussions. The S10 data are available in http://sol.spacenvironment.net/jb2008/indices/SOLFSMY.TXT. We acknowledge extensive use of NASA's Astrophysics Data System.